# Inferring statistics of planet populations by means of automated microlensing searches

**M. Dominik[1⋆†], U. G. Jørgensen[2], K. Horne[1], Y. Tsapras[3], R. A. Street[4], Ł. Wyrzykowski[5,6], F. V. Hessman[7], M. Hundertmark[7], S. Rahvar[8], J. Wambsganss[9], G. Scarpetta[10,11], V. Bozza[10,11], S. Calchi Novati[10,11], L. Mancini[10,11], G. Masi[12], J. Teuber[2], T. C. Hinse[2,13], I. A. Steele[14], M. J. Burgdorf[14], S. Kane[15]**

[1]SUPA, University of St Andrews, School of Physics & Astronomy, North Haugh, St Andrews, KY16 9SS, United Kingdom
[2]Niels Bohr Institutet, Københavns Universitet, Juliane Maries Vej 30, 2100 København Ø, Denmark
[3]Las Cumbres Observatory Global Telescope Network, 1 Morpeh Wharf, Birkenhead, Merseyside, CH41 1NQ, United Kingdom
[4]Las Cumbres Observatory Global Telescope Network, 6740B Cortona Dr, Goleta, CA 93117, United States of America
[5]Institute of Astronomy, University of Cambridge, Madingley Road, Cambridge CB3 0HA, United Kingdom
[6]Warsaw University Astronomical Observatory, Al. Ujazdowskie 4, 00-478 Warszawa, Poland
[7]Institut für Astrophysik, Georg-August-Universität, Friedrich-Hund-Platz 1, 37077 Göttingen, Germany
[8]Department of Physics, Sharif University of Technology, P. O. Box 11365–9161, Tehran, Iran
[9]Zentrum für Astronomie der Universität Heidelberg (ZAH), Astronomisches Rechen-Institut, Mönchhofstr. 12-14, 69120 Heidelberg, Germany
[10]Università degli Studi di Salerno, Dipartimento di Fisica "E.R. Caianiello", Via S. Allende, 84081 Baronissi (SA), Italy
[11]INFN, Gruppo Collegato di Salerno, Sezione di Napoli, Italy
[12]Bellatrix Astronomical Observatory, Via Madonna de Loco 47, 03023 Ceccano (FR), Italy
[13]Armagh Observatory, College Hill, Armagh, BT61 9DG, Northern Ireland
[14]Astrophysics Research Institute, Liverpool John Moores University, Twelve Quays House, Egerton Wharf, Birkenhead, CH41 1LD, United Kingdom
[15]NASA Exoplanet Science Institute, Caltech, MS 100-22, 770 South Wilson Avenue, Pasadena, CA 91125, United States of America

*ABSTRACT*

The study of other worlds is key to understanding our own, and by addressing formation and habitability of planets, one not only investigates the origin of our civilization, but also looks into its future. With a bunch of extraordinary characteristics, gravitational microlensing is quite a distinct technique for detecting and studying extra-solar planets. Rather than in identifying nearby systems and learning about their individual properties, its main value is in obtaining the statistics of planetary populations within the Milky Way and beyond. Only the complementarity of different techniques currently employed promises to yield a complete picture of planet formation that has sufficient predictive power to let us understand how habitable worlds like ours evolve, and how abundant such systems are in the Universe. A cooperative three-step strategy of survey, follow-up, and anomaly monitoring of microlensing targets, realized by means of an automated expert system and a network of ground-based telescopes is ready right now to be used to obtain a first census of cool planets with masses reaching even below that of Earth orbiting K and M dwarfs in two distinct stellar populations, namely the Galactic bulge and disk. In order to keep track with the vast data volume that needs to be dealt with in real time in order to fulfill the science goals, and to allow the proper extraction of the planet population statistics, fully-automated systems are to replace human operations and decisions, so that the hunt for extra-solar planets thereby acts as a principal science driver for time-domain astronomy with robotic-telescope networks adopting fully-automated strategies. Several initiatives, both into facilities as well as into advanced software and strategies, are supposed to see the capabilities of gravitational microlensing programmes step-wise increasing over the next 10 years. New opportunities will show up with high-precision astrometry becoming available and studying the abundance of planets around stars in neighbouring galaxies becoming possible. Finally, with the detection of extra-solar planets (not only by gravitational microlensing) and the search for extra-terrestrial life being quite popular topics already, we should not miss out on sharing the vision with the general public, and make its realization to profit not only the scientists but all the wider society.

## 1 INTRODUCTION

Gazing at the bright little spots on the night sky has probably ever inspired human mind to wonder whether there are other worlds like ours, and to reflect about our own existence. While it was anticipated for a long time that planets around other stars do exist, the first such was not detected before 1995 (Mayor & Queloz 1995). Within less than 15 years, more than 300 extra-solar planets were revealed,[1] and the study of the distribution of their orbital properties, substantially different from those of the planets in the Solar system, has revolutionized our understanding of how planets form and planetary systems evolve. While the vast majority of the discovered extra-solar planets are gas giants, similar to Jupiter or Saturn, one meanwhile knows about 10 planets below 10 Earth masses, so-called 'Super-Earths', which are thought to come in a variety of flavours, depending on whether they are primarily made of iron, silicates, water, or carbon compounds (e.g. Seager et al. 2007).

Ultimately, we would like to understand the origin of planet Earth and how our civilization emerged. It is in fact the study of other worlds that holds a key for understanding our own. Apart from the question about the origin of mankind and its role in the Universe, such studies also examine the border between habitability and inhabitability and thereby touch (in the primary sense of that word) vital issues about our ecosphere, where some of them are likely to contribute to the topical debate about climate change and greenhouse gases (e.g. Strelkov 1966; Kasting 1988; Sidorov & Parot'kin 1991), identified to be crucial for our future.

The different indirect techniques for detecting exoplanets that have so far proven successful, namely relying on Doppler-wobbles (Mayor & Queloz 1995), planetary transits (Henry et al. 2000; Udalski et al. 2002), or gravitational microlensing (Bond et al. 2004), have rather different characteristics which leads to them occupying (partly overlapping) complementary regions of planet discovery space, as schematically illustrated in Fig. 1. Efforts are underway for all of these techniques to push towards planets more closely resembling Earth, but it is important to be aware of the fact that habitable systems can only be understood once models of planet formation and orbital migration can be made to match observations over an embracing wider region of parameter space. Moreover, the abundance of planets suitable to be populated by life forms as we know them on Earth can only reliably be estimated if we know how typical or special the Solar system is, and how frequently planets form around stars of different type and population.

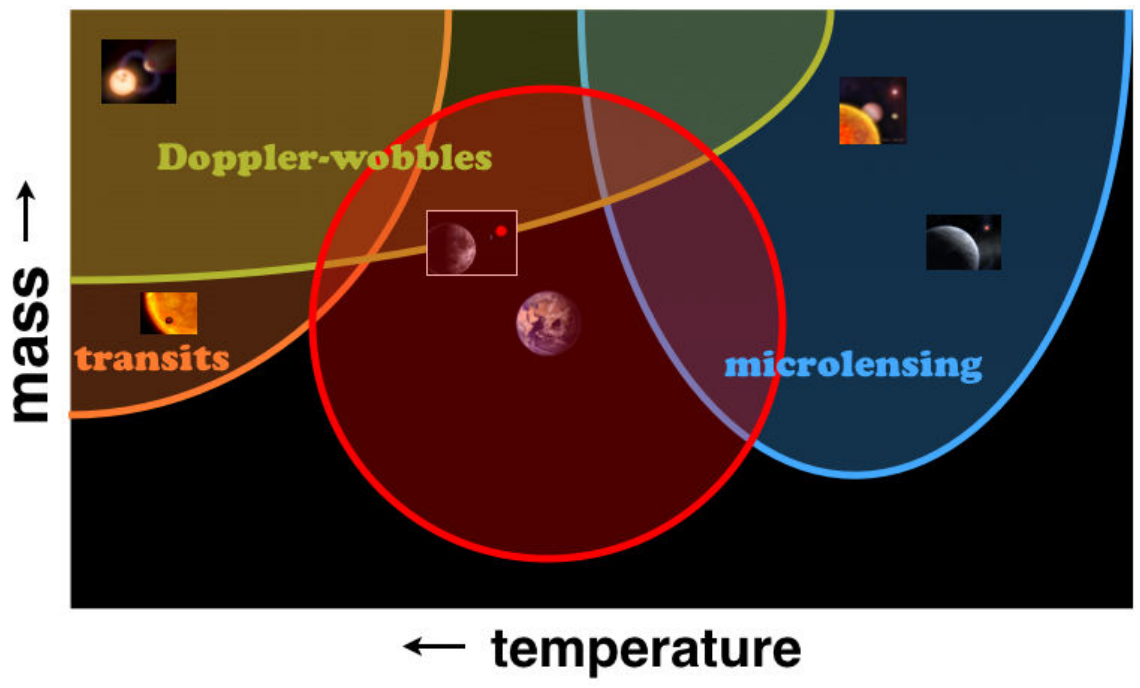

**Figure 1.** Schematic view of the characteristic sensitivities as a function of planet mass and temperature for the three indirect techniques that have so far proven successful for detecting extra-solar planets; illustrating how planets similar to Earth are being approached, and an embracing region (indicated in red) giving power to probe planet-formation models is covered.

⋆ Royal Society University Research Fellow
† E-mail: `md35@st-andrews.ac.uk`
[1] `http://wwww.exoplanet.eu`

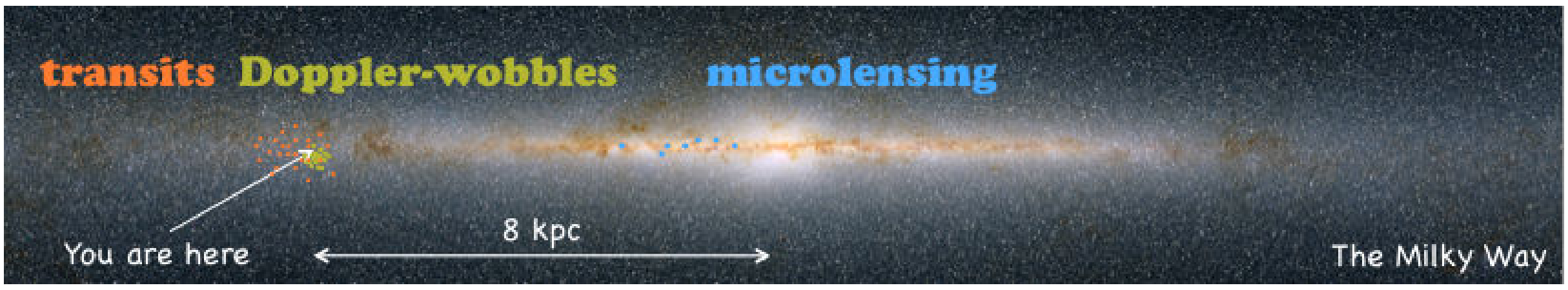

**Figure 2.** Locations of the detected planets within the Milky Way for each of the three successful indirect detection techniques applied from the ground. Contrary to the other approaches, gravitational microlensing allows to obtain a census of planets orbiting either Galactic disk or bulge stars, thereby collecting information about two distinct stellar populations, for which planet formation might differ. The only other known (complementary) approach to study planets around Galactic bulge stars is a space-based transit survey (Sahu et al. 2006), sensitive to hot rather than cool planets.

The effect of gravitational microlensing provides a unique opportunity to obtain a census of cool planets, including not only super-Earths, but even sub-Earths, as well as planetary systems like our own for two different stellar populations, namely the Galactic disk and bulge, rather than being limited to closeby objects. Fig. 2 explicitly shows the locations of the detected planets within the Milky Way for each of the three techniques.

Nearly all of the planets being detected by microlensing reside beyond the cold edge of the habitable zone (Kasting et al. 1993), and the overwhelming majority are found to orbit K- and M-dwarf stars. However, for (less abundant) more massive host stars there is some overlap with the habitable zone, constituting $\lesssim 3\,\%$ of the expected detections (Park et al. 2006). Planets revealed from a microlensing signature are too distant for any follow-up observations, and the microlensing technique only gives us a short non-repeating window-of-opportunity (lasting between a few hours and a few days depending on the mass of the planet), although some information about its host star can be obtained later. Nevertheless, their census provides insight into our perspective for life within the Milky Way and beyond, that is not obtainable by other means.

With the realization of mankind's vision to detect signatures of life on planets around stars other than the Sun getting into reach, it is understandable that before committing to invest several billion euros into large cutting-edge space projects, one would like to know whether the objects that are to be studied are abundant enough for such missions becoming a likely success.

The discovery of the icy planet OGLE-2005-BLG-390Lb (Beaulieu et al. 2006) by microlensing has not only impressively demonstrated the capability of the applied technique in the super-Earth mass regime, but more importantly provided a first observational hint that such planets are common in the Universe. In addition, the double catch of two gas-giant planets orbiting OGLE-2006-BLG-109L (Gaudi et al. 2008), constituting a 'look-alike' of the Solar System – with the respective planets matching the hierarchy of Jupiter and Saturn with respect to the snow line – indicates that the Solar System is not a rare type amongst planetary systems, but instead those known so far constitute a strongly-biased sample, and moreover that planetary systems are the rule rather than the exception.

With the microlensing planet detections that have been claimed so far appearing concentrated around massive gas-giants or super-Earths, respectively, there might be a hint at a planet mass gap for M dwarfs, but this is not statistically significant (yet). Currently favoured models of planet formation based on core accretion (Ida & Lin 2005) reasonably assume that the surface density of a protoplanetary disk around M dwarfs is much smaller than around Solar-type stars. As a consequence, mass accretion is slow, and planets will not be able to accrete all the solid material in their respective feeding zone for outer regions at $\gtrsim$ few AU. The resulting cut-off of the planetary population, which could (now and in the foreseeable future) uniquely be identified by the statistics of microlensing observations, thereby measures the disk surface density around M dwarfs as well as the accretion rate of planetesimals, which constitute fundamental and crucial parameters for the underlying theories.

## 2 PLANET DETECTION BY MICROLENSING

(Galactic) *gravitational microlensing*[2] is understood as the transient brightening of an observed star caused by the gravitational bending of light due to an intervening foreground star. For a background *source star* at distance $D_\mathrm{S}$ and a foreground *lens star* with mass $M$ at distance $D_\mathrm{L}$, the unique characteristic scale of gravitational microlensing is given by the *angular Einstein radius* (Einstein 1936)

$$\theta_\mathrm{E} = \sqrt{\frac{4GM}{c^2}\,\left(D_\mathrm{L}^{-1} - D_\mathrm{S}^{-1}\right)}\,, \tag{1}$$

while a source star separated by an angle $u\,\theta_\mathrm{E}$ from the lens star is magnified by

$$A(u) = \frac{u^2+2}{u\,\sqrt{u^2+4}}\,, \tag{2}$$

so that $\theta_\mathrm{E}$ quantifies the angular alignment required to create an observable effect. For source stars in the Galactic bulge, typical distances of $D_\mathrm{S} = 8.5$ kpc and $D_\mathrm{L} = 6.5$ kpc yield $\theta_\mathrm{E} \sim 540\ \mu\mathrm{as}\,(M/M_\odot)^{1/2}$. If one assumes a uniform proper motion $\mu$ between lens and source star, one finds

$$u(t) = \sqrt{u_0^2 + \left(\frac{t-t_0}{t_\mathrm{E}}\right)^2}\,, \tag{3}$$

so that microlensing light curves are symmetric with respect to the epoch $t_0$, for which a peak magnification $A_0 = A(u_0)$ is reached, while the duration of the event is characterized by the event time-scale $t_\mathrm{E} \equiv \theta_\mathrm{E}/\mu$ (Paczyński 1986). With the kinematics of the Milky Way yielding $\mu \sim 15\ \mu\mathrm{as\,d}^{-1}$, microlensing events typically last $t_\mathrm{E} \sim 35\ \mathrm{d}\,(M/M_\odot)^{1/2}$. Moreover, as

[2] as opposed to cosmological gravitational microlensing of observed quasars

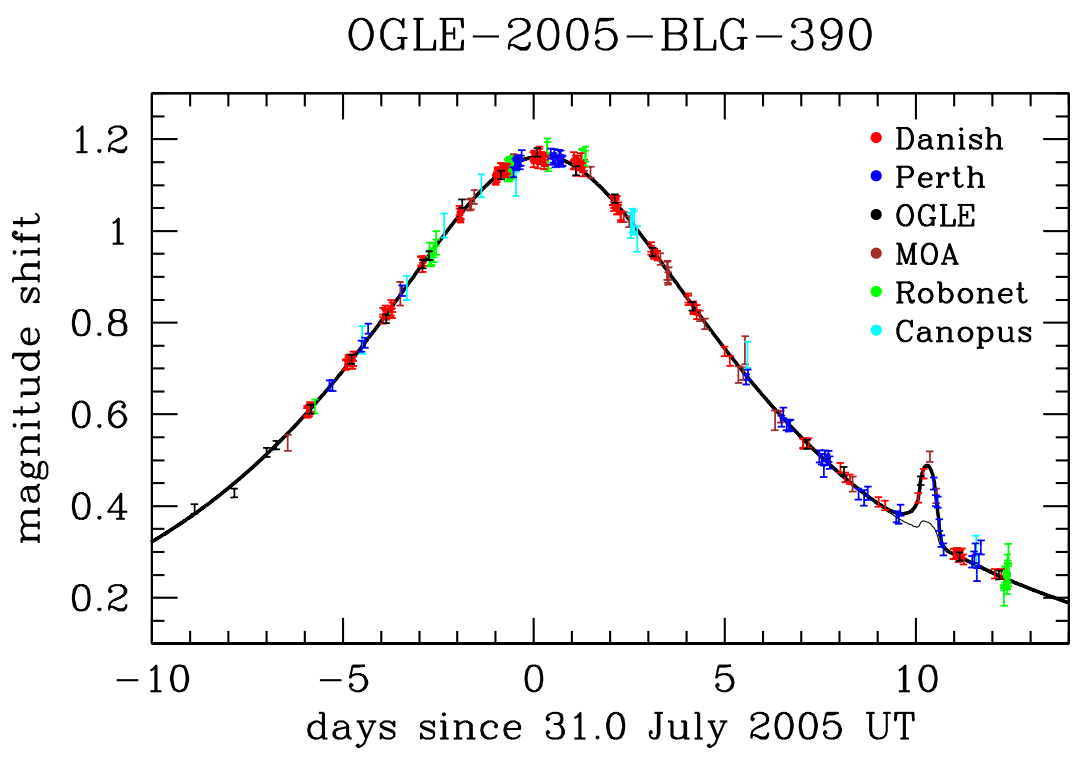

**Figure 3.** Model light curve, showing the magnitude shift, i.e $2.5\ \lg A[u(t)]$, where $A$ denotes the magnification of the observed source star (see Eq. (2)), for microlensing event OGLE-2005-BLG-390 along with the data collected with 6 different telescopes (colour-coded). The $\sim 15\,\%$ blip, lasting about a day, revealed the 5-Earth-mass planet OGLE-2005-BLG-390Lb (Beaulieu et al. 2006). The thinner line refers to the hypothetical detectable $3\,\%$ deviation that would have arisen from an Earth-mass planet in the same spot.

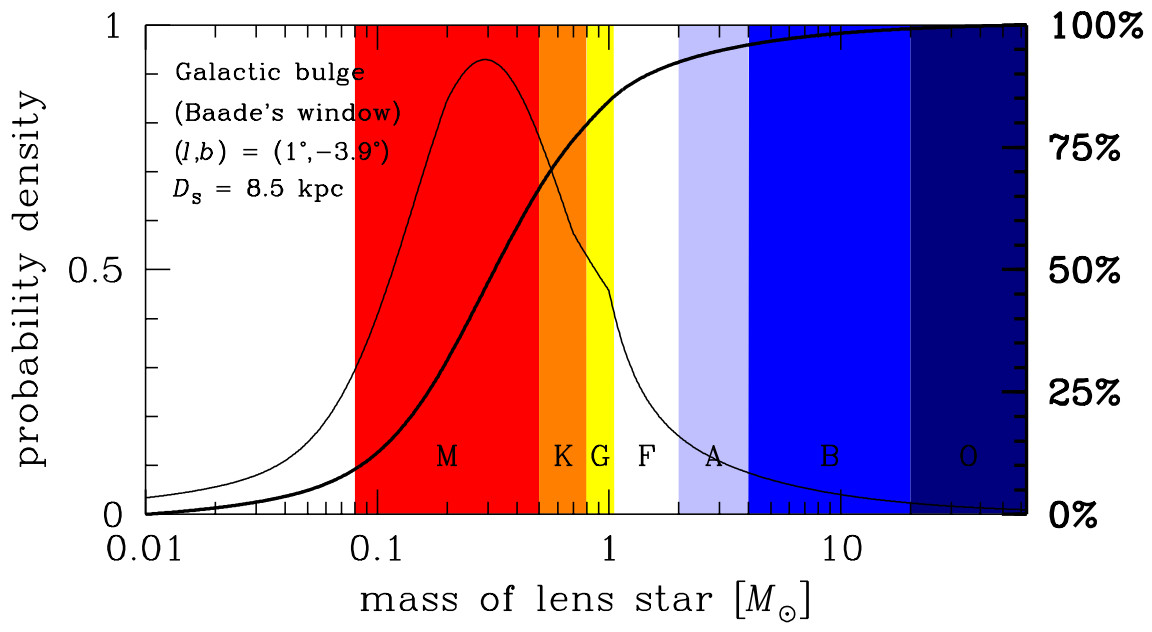

**Figure 4.** Distribution of the lens stars causing microlensing events within the Milky Way with their mass and corresponding spectral type. These constitute the hosts for planets that can be revealed. The source star is considered to be located in the Galactic bulge towards Baade's window at galactic coordinates $(l, b) = (1°, -3.9°)$ and a distance $D_\mathrm{S} = 8.5$ kpc. A model of the Milky Way consisting of a tilted barred bulge and a double-exponential disk has been assumed following the choice of Dominik (2006), while bulge and disk stellar mass functions suggested by Chabrier (2003) have been adopted. The thick line corresponds to the cumulative distribution (scale on right axis), whereas the thin line refers to the probability density (scale on left axis). The distribution shown refers to the ongoing events and not the observed ones, for which there are selection effects, e.g. related to the event time-scale or blending of the observed target by the lens star.

long as the brightness profile of the source star can be neglected, microlensing events are perfectly achromatic.

As first pointed out by Mao & Paczyński (1991), planets orbiting the foreground lens star (rather than the observed source star) can reveal their existence by creating deviations to the observed light curves by means of their mass altering the gravitational bending of light, such as the $\sim 15\,\%$ signature of the 5-Earth-mass planet OGLE-2005-BLG-390Lb (Beaulieu et al. 2006) shown in Fig. 3.

If one defines a microlensing event to be ongoing while $u < 1$, corresponding to magnifications $A > 3/\sqrt{5} \approx 1.34$, the event rate $\Gamma$ becomes the product of the number of observed source stars $N_\mathrm{S}$, the number area density of the lens stars, the effective transverse velocity $v = D_\mathrm{L}\,\mu = |\boldsymbol{v}_\mathrm{L} - x\,\boldsymbol{v}_\mathrm{S} - (1-x)\,\boldsymbol{v}_\mathrm{O}|$ – where $\boldsymbol{v}_\mathrm{L}$, $\boldsymbol{v}_\mathrm{S}$, and $\boldsymbol{v}_\mathrm{O}$ are the velocities of the lens, source, and observer, respectively, perpendicular to the line-of-sight, and $x = D_\mathrm{L}/D_\mathrm{S}$ –, and the cross-section $2\,D_\mathrm{L}\,\theta_\mathrm{E}$, so that

$$\Gamma = \frac{4\sqrt{G}}{c}\,N_\mathrm{S}\,D_\mathrm{S}^{3/2}\int_0^1 \rho(xD_\mathrm{S})\,\sqrt{x(1-x)}\,v(x)\,\mathrm{d}x\ \times \times \int_0^\infty \sqrt{M}\,\xi(M)\,\mathrm{d}M\,, \tag{4}$$

with $\rho(D_\mathrm{L})$ denoting the volume mass density and $\xi(M)$ the mass function of the lens stars, where it has been assumed that the latter does not depend on $D_\mathrm{L}$. This expression reveals that one should aim at monitoring as many source stars as possible, which should be as distant as possible (however without compromising on the photometric accuracy), both because the event rate increases with distance $\Gamma \propto D_\mathrm{S}^{3/2}$ as well as with the amount of intervening matter. The lens stars, hosting the detectable planets, will most likely be located in dense regions, with further preference being given to distances about half-way between observer and source star. Microlensing effects are more likely to involve more massive ($\Gamma \propto \sqrt{M}$), but more importantly more abundant host stars. As Fig. 4 shows, this makes M dwarfs the most prominent choice. Moreover, such low-mass stars are further preferred because the larger amount of light emitted by more massive lens stars substantially blends the observation of main-sequence source stars, thereby reducing the observed magnification during the course of the event.

The detection of planets by microlensing is aided by a 'resonance', which increases the detection probability (Mao & Paczyński 1991; Gould & Loeb 1992), occuring if the angular separation $\theta_\mathrm{p}$ of the planet from its host star happens to match the angular Einstein radius $\theta_\mathrm{E}$, corresponding to a separation parameter $d \equiv \theta_\mathrm{p}/\theta_\mathrm{E} = 1$. It is a lucky coincidence that the gravitational radius of stars $R_\mathrm{S} = (2GM)/c^2$ of a few km and the effective distance $D = (D_\mathrm{L}^{-1} - D_\mathrm{S}^{-1})^{-1}$ of several kpc combine to a separation $r_\mathrm{E} = D_\mathrm{L}\,\theta_\mathrm{E}$ of a few AU at the distance of the planet's host star.

The tidal gravitational field of the lens star at the position of the planet breaks the local symmetry and thereby creates *planetary caustics* (Chang & Refsdal 1979). Moreover, an extended *central caustic* arises near the position of the lens star as a result from the symmetry-breaking by all orbiting planets. While planetary caustics are practically always associated with a single planet and interactions between the planets are extremely unlikely (Bozza 1999, 2000), the central caustic is always the result of all planets that orbit the respective star (Gaudi, Naber & Sackett 1998). Whenever the source star gets aligned with a caustic, its magnification becomes quite large (and approaches infinity as its angular size tends to zero). This behaviour is reflected in the fractional change to the magnification caused by a planet as function of the angular position of the source star relative to the lens star and its planet, as illustrated in Fig. 5.

The caustic topology matches one of three types (Schneider & Weiß 1986; Erdl & Schneider 1993), so that all star-planet systems can be sorted into a respective category according to the planet-star separation: close, intermediate, and wide (Dominik 1999b). Around the 'resonant' case, $d = 1$, one finds a single intermediate caustic (a hybrid of the planetary and central caustic), where this region narrows down towards smaller mass ratios, whereas there are two separate caustics for the wide-binary case (diamond-shaped central and planetary caustics), and three caustics for the close-binary case (a diamond-shaped central caustic and two triangular-shaped planetary caustics). The planetary caustics are within the Einstein circle (of radius $\theta_\mathrm{E}^{[1]}$ around the lens star)[3] for a planet in the *lensing zone* $(\sqrt{5}-1)/2 \leqslant d^{[1]} \leqslant (\sqrt{5}+1)/2$,[4] where $d^{[1]} \equiv \theta_\mathrm{p}/\theta_\mathrm{E}^{[1]}$, which gives a range of separations with a large probability to reveal it. Obviously, larger planet-to-star mass ratios $q$ provide larger regions of the sky for the source star to exhibit a detectable change in magnification of a given amount due to the planet. Comparing their sizes as a function of $q$, one finds that their linear extent roughly drops $\propto \sqrt{q}$ near the planetary caustics and $\propto q$ near the central caustic. Given that the observed light curves correspond to one-dimensional cuts through the excess magnification maps as shown in Fig. 5, both the duration $t_p$ of the planetary signal and the probability for it to occur during the course of a microlensing event are proportional to the same respective factor ($\sqrt{q}$ or $q$), while signals of any given amplitude could in principle arise for arbitrarily small mass ratios, as long as the source star can be fairly approximated as point-like. The finite angular radius $\theta_\star$ of the source star however smears out the planetary signal, thereby reducing its amplitude while increasing its duration, where $t_p \sim 2\,\theta_\star/\mu$ becomes related to the size of the source star rather than the mass of the planet. As long as the signal amplitude does not fall below the detection threshold, this can actually increase the planet detection efficiency (as for OGLE-2005-BLG-390Lb), but otherwise it limits the detectability of planets towards small masses, which affects sources near the central caustic far more seriously than near planetary caustics due to its smaller size.

[3] The superscript [1] stresses the strict reference to the mass of the star only, i.e the primary object.

[4] These values are related to the golden ratio.

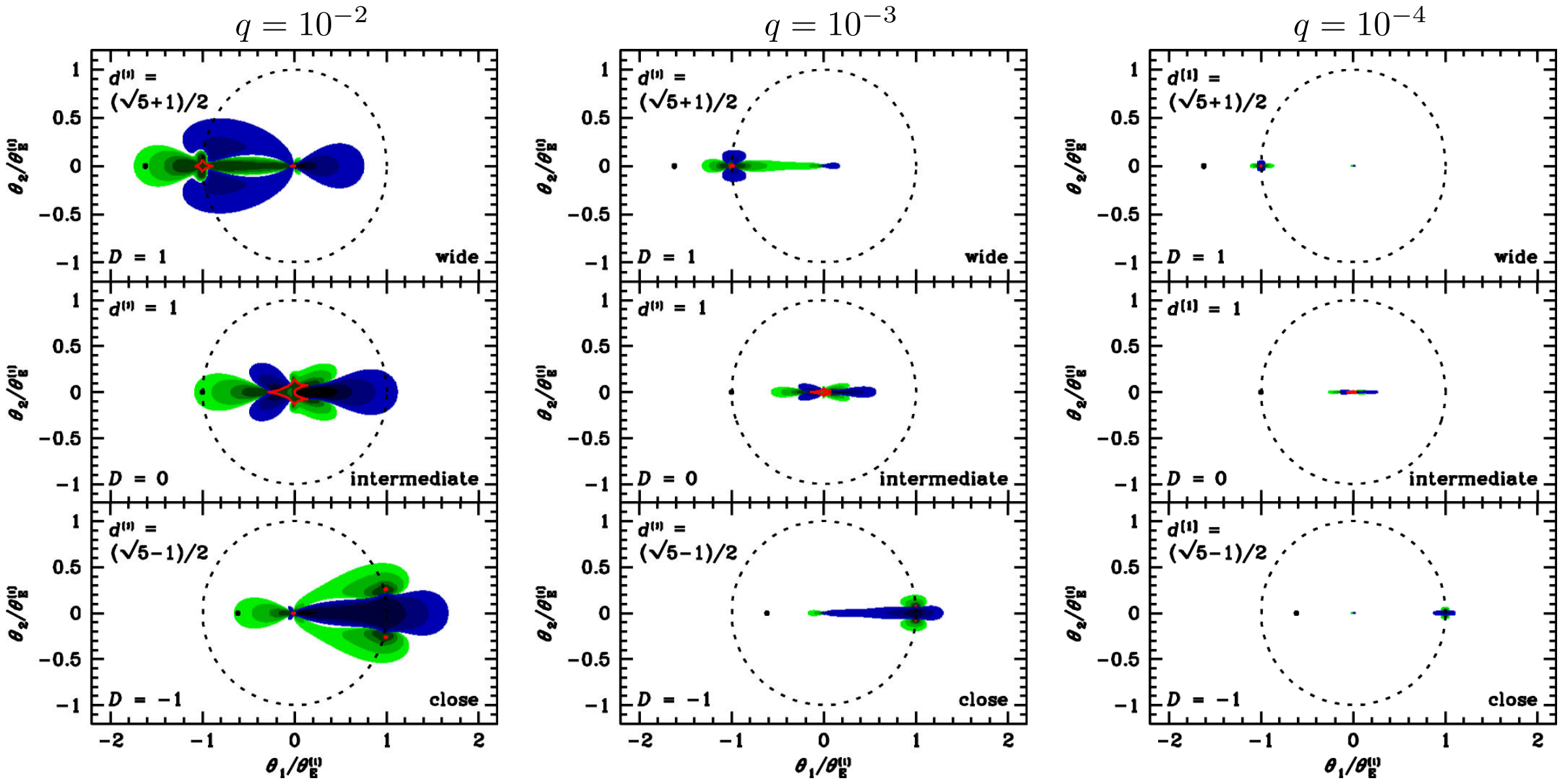

**Figure 5.** Locations for the source star on the sky where a planet orbiting the lens star would create a significant difference in magnification. While green shades correspond to a further brightening in presence of the planet, blue shades correspond to a dimming, where the shade levels refer to fractional excess magnifications of 1 %, 2 %, 5 % and 10 %. For comparison, the caustics are shown in red. From left to right, three different mass ratios have been chosen, namely $q = 10^{-2}$, $q = 10^{-3}$, and $q = 10^{-4}$, roughly corresponding to planets of 3 Jupiter masses, Saturn mass, and 10 Earth masses, respectively, for a stellar mass of $0.3\ M_{\odot}$. The star is located in the centre of the coordinate system and coordinates refer to angles in units of the angular Einstein radius $\theta_E^{[1]}$ of the stellar mass. For each of the shown panels, the planet is located along the horizontal axis at a separation $\theta_p$ from its host star, where this position is marked by a black filled circle and $d^{[1]} \equiv \theta_p/\theta_E^{[1]}$. For each mass ratio, three different separations have been chosen, corresponding to the two boundaries of the 'lensing zone' and the resonant case where the planet happens to lie on the Einstein circle of radius $\theta_E^{[1]}$ (shown as dashed line) of the lens star, which illustrate the three different caustic topologies. $D = d^{[1]} - (d^{[1]})^{-1}$ marks the position of the planetary caustics.

With the time-scale of the planetary signal being related to the motion of stars within the Milky Way, their distances, the mass of the planet, and/or the angular size of the observed source stars, but *not* to the orbital period of the planet, the orbital period does not place a limit on the detection, contrary to all other proposed indirect techniques. This allows planets in wide orbits to be found, whose periods would otherwise exceed the lifetime of experiments or their investigators.

## 3 OBSERVING STRATEGY AND REQUIREMENTS

The detection of planets by microlensing is mainly a matter of probability and statistics, rather than a matter of principle. In order to provide a large enough sample, one would like to monitor as many suitable targets as possible. The crowding level of the observed fields and the resolution of the detector are important issues, since stars that are blurred into a single pixel (Baillon et al. 1993) challenge the efforts for measuring the brightness variation at the desired accuracy. Given that the field-of-view of any instrument is limited, the capabilities are determined by the exposure time required to achieve a photometric accuracy that allows to fulfill the scientific aims.

With its stellar density, the Galactic bulge is a favourable direction (Kiraga & Paczyński 1994), albeit that fainter objects are likely to be blended with nearby brighter stars (e.g. Woźniak & Paczyński 1997; Vermaak 2000; Smith et al. 2007), and much of the central parts of the Milky Way are not visible at optical wavelengths.

A high-cadence wide-field survey to hunt for Earth-mass planets has already been discussed by Sackett (1997), and specific prospects have more recently been investigated by Han (2007b). However, a realization so far suffered from the lack of available facilities that can be committed to such a large programme. This approach might however become an option in about five years' time. A more ambitious microlensing space mission has also been advocated (Bennett & Rhie 2002; Bennett et al. 2003), realizable in about 10 years, which would profit from improved photometry and the elimination of gaps in the coverage of events due to bad weather. On the other hand, as all space missions, it needs to compete against the advance of ground-based technology over its rather long development phase, and has the disadvantage of being a one-off project with strictly limited life-time. Given the available facilities, a cooperative survey/follow-up concept has evolved from the need to find a compromise between the field-of-view, the sampling rate, the limiting magnitude, and the resolution of targets. Such a strategy has been adopted as early as 1995 by the PLANET (*P*robing *L*ensing *A*nomalies *NET*work) collaboration (Albrow et al. 1998; Dominik et al. 2002) in conjunction with the OGLE (*O*ptical *G*ravitational *L*ensing *E*xperiment) and MACHO (*MA*ssive *C*ompact *H*alo *O*bject) projects (Udalski et al. 1992; Alcock et al. 1995a), and has been followed by several further teams since then, with step-wise increased capabilities. While the survey telescopes monitor the stellar fields in the Galactic Bulge for microlensing events on a daily basis, the follow-up sites observe just a most promising fraction of the ongoing events at a higher sampling rate in order to detect planetary signals or prove their absence.

The selection of further details of a most powerful strategy is strongly dependent on the availability of telescopes that can be committed to such a science programme. As discussed in the previous section and illustrated in Fig. 5, planets reveal their existence whenever the source star enters the vicinity of either the central or planetary caustics, provided that sufficient data are collected over the respective epochs. If the lens star happens to pass very closely to the line-of-sight to the observed source star, one probes central caustics and their surroundings, where potential deviations are found around a highly-magnified peak, which can reliably be predicted about 12–36 hrs in advance (Albrow 2004). Since a large range of possible (undetermined) orientation angles of the relative motion between source and lens star on the sky produces a detectable signal, the efficiency for detecting planets approaches a lensing-zone average of 100 % for gas giants if peak magnifications $A_0 \gtrsim 10$ arise, and it is still quite substantial for planets of Neptune mass (Griest & Safizadeh 1998; Rattenbury et al. 2002). However, among all ongoing events, those with smaller impact parameter $u_0$ (as defined by Eq. (3)), and therefore larger peak magnification $A_0$, become less and less abundant the larger the considered threshold in $A_0$ (and the smaller that in $u_0$) gets. In particular towards less massive planets, the detection from approaches to planetary caustics harbours the larger total potential, given the larger associated deviation regions that are less susceptible to a washout by finite-source effects. However, making use of these enlarged theoretical prospects requires a much larger effort, which involves the monitoring of lots of events, each having a more moderate detection efficiency, namely $\sim 20\,\%$ for jupiter-mass planets in the lensing zone amongst all events with $A_0 > 1.34$ (Gould & Loeb 1992), over a substantial fraction of their duration, since the deviation could occur anytime.

Efforts on just monitoring peaks of events at high magnification are well-suited to yield spectacular results on individual events that can have immediate consequences, such as the double catch of gas-giant planets orbiting OGLE-2006-BLG-109L for the uniqueness of the Solar system (Gaudi et al. 2008), but these cannot provide the statistics about extra-solar planets resulting from the more demanding regular monitoring by a dedicated network of telescopes. Despite the fact that events with extremely small impact parameters, such as OGLE-2004-BLG-343 (Dong et al. 2006), where $u_0 \lesssim 5 \times 10^{-4}$, provide an excellent and low-effort opportunity to study not only planets below Neptune mass, but the complete planetary system (Gaudi et al. 1998), these are far too rare for providing statistically significant results. In fact, events as promising as OGLE-2004-BLG-343 or better can be expected to be spotted less than once per 5 years with current surveys. Moreover, the extraordinary high detection efficiency is limited to a rather narrow range of orbital separations around $d \sim 1$, so that only a small fraction of the planet population is probed.

It was not a surprise that the approach of regular dense monitoring of a as large as possible number of events was successful in revealing the 5-Earth-mass planet OGLE-2006-BLG-390Lb (Beaulieu et al. 2006). As Fig. 3 shows, an Earth-mass planet in the same spot would have provided a 3 % deviation lasting about 12 hrs. It could have been detected if the regular dense monitoring (of one to two hours) had been intensified once a planetary deviation is suspected to be in progress. This makes a strong case for extending the two-step survey/follow-up strategy to a three-step strategy of survey, follow-up, and anomaly monitoring. Such a suggestion had in fact already entered NASA's ExNPS (*Ex*ploration of *N*eighboring *P*lanetary *S*ystems) Roadmap (Elachi et al. 1996), but it was not systematically realized for the subsequent ten years; which could have been possible, but failed due to insufficient funds being made available. While several teams have tried to communicate potential deviations in progress by means of human monitoring and decision-making, the adopted procedures have shown not to be very efficient for detecting planets, and have failed in prominent cases. In particular, the OGLE observations on event OGLE-2002-BLG-055 show a single deviant point, giving no rise to the suspicion that something is wrong with its photometry, and the complete light curve is perfectly compatible with a planet orbiting the lens star (Jaroszyński & Paczyński 2002). However, the lack of coverage prevents claiming a detection, and there are several alternative explanations for the nature of this event (Gaudi & Han 2004). This prompted for the implementation of OGLE's EEWS (*E*arly *E*arly *W*arning *S*ystem), flagging deviant data immediately to the observer (Udalski 2003). A modified approach is required to account for robotic telescopes (where no observer is present) and multi-site observations, as it is being achieved by the SIGNALMEN anomaly detector (Dominik et al. 2007).The automated detection of suspected anomalies becomes an indispensable requirement for a powerful strategy, given that one needs to monitor lots of events, needs to decide promptly, and needs to communicate the results immediately (which means within 5–10 minutes). Fig. 6 shows how the additional step of anomaly monitoring following an automated trigger pushes the detection efficiency towards planets that are three times less massive than those detectable from a pure survey/follow-up scheme – and Earth mass turns out not to be the limit for current microlensing planet searches.

## 4 INFERRING THE PLANET CENSUS

The technique of gravitational microlensing provides detections of planets that are not observed, orbiting stars that are not observed either. Moreover, their signature will only show once, so that the "detections" do not fulfill the scientific requirement of repeatability and the opportunity of independent confirmation or falsification.[5] This implies that one cannot, in contrast to repeatable studies, move towards a larger degree of belief approaching certainty about a detection by means of further observations on the same object. Similar to the capabilities of the technique of gravitational microlensing being mostly of statistical nature, it is again the statistics that matter. In fact, the statistics of planet detections by microlensing do repeat (probabilistically) and will converge towards a consistent picture.

The observed planetary deviations in microlensing light curves are a statistical representation of the convolution of the underlying distribution function of the properties of the planets as realized in nature and the detection efficiency of the experiment. Therefore, in order to be able to learn about the planetary population, one needs both to determine the detection efficiency and to assess the probability for planets within a given range of parameters to reside around the host star of the respective observed microlensing event. While the quantity of information gained is a function of the detection efficiency, it does not depend on whether planets are actually revealed or not (although the quality of information is different). In fact, the dense regular monitoring of microlensing events provided valuable constraints on the planetary abundance around M dwarfs (Albrow et al. 2001; Gaudi et al. 2002; Tsapras et al. 2003; Snodgrass et al. 2004) much before the first planet detection (Bond et al. 2004) could have been reported. Therefore, one might well consider to adopt a strategy that maximizes the planet detection efficiency (which does not depend on the actual abundance of planets within a certain range of parameters) rather than the number of detections (which obviously does).

[5] However, there can well be independent evidence if planetary signatures happen to be monitored by several telescopes while those are lasting.

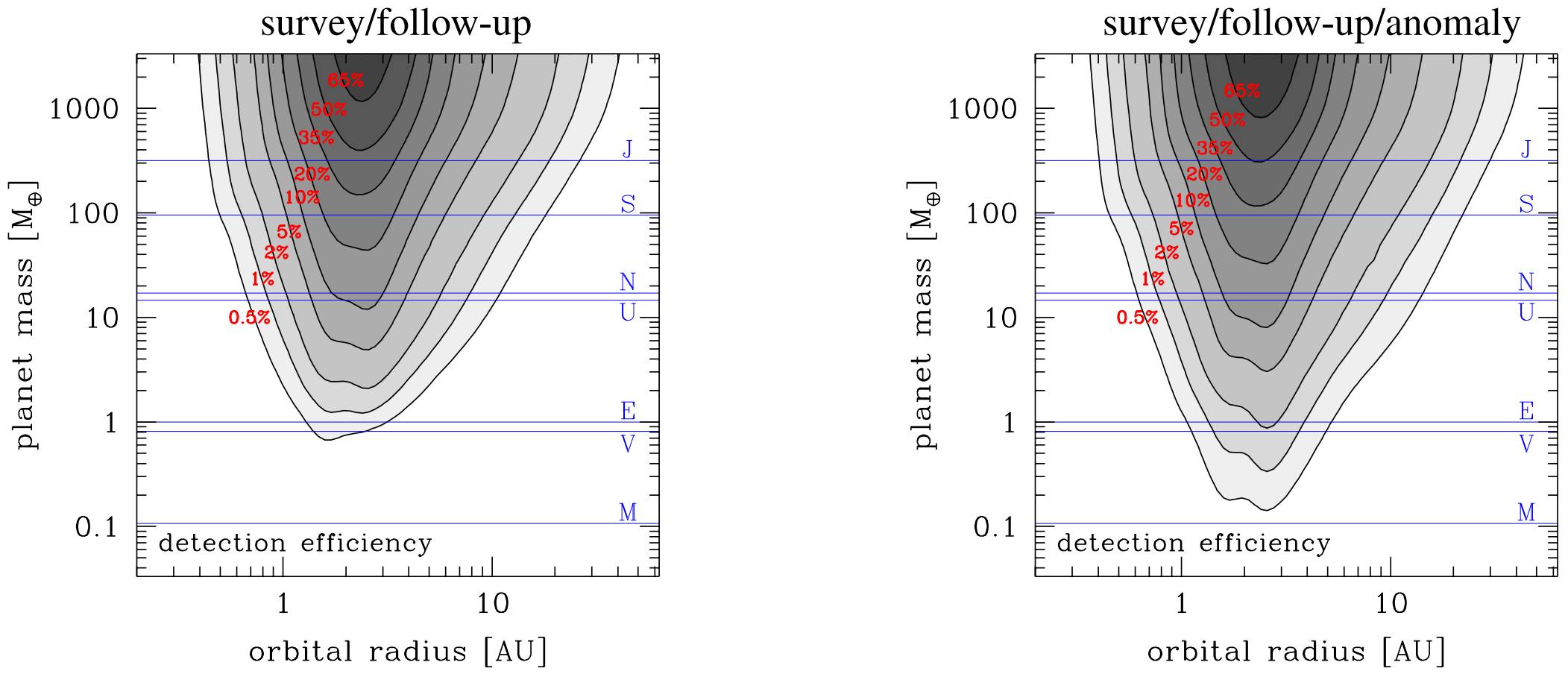

**Figure 6.** The planet detection efficiency as a function of the orbital radius (assuming a circular orbit) and the mass of the planet for a survey/follow-up microlensing planet search, and how high-cadence sampling triggered by an automated anomaly detector pushes it to significant values for planet masses even below that of Earth. A typical microlensing event with angular Einstein radius $\theta_\mathrm{E} = 274$ µas and proper motion $\mu = 13.7$ µas d$^{-1}$, yielding an event time-scale $t_\mathrm{E} \equiv \theta_\mathrm{E}/\mu \sim 20$ d, has been assumed. The adopted source-size parameter $\rho_\star \equiv \theta_\star/\theta_\mathrm{E} = 0.002$ corresponds to a main-sequence star with radius $R_* = D_\mathrm{S}\,\theta_\star \sim 1\,R_\odot$ in the Galactic bulge at a distance $D_\mathrm{S} \sim 8.5$ kpc. The closest angular approach between lens and source star of $0.3\,\theta_\mathrm{E}$ results in a peak magnification $A_0 \sim 3.5$. A planet is considered to be "detected" by observing a sequence of consecutive points on the same side of the model light curve of which at least 5 deviate by more than twice their photometric uncertainty, assumed to be 3 % at baseline with a further systematic error of 0.5 % added in quadrature. For the sampling intervals, $\Delta t = 1$ d has been chosen for the surveys, $\Delta t = (90\ \mathrm{min})\ A^{-1/2}$ for follow-up observations, starting at a magnification $A = 1.5$ and stopping at $A = 1.06$, and $\Delta t = 5$ min for anomaly monitoring, activated after a data point is found to deviate by more than $2\,\sigma$.

The planet detections claimed so far (Bond et al. 2004; Udalski et al. 2005; Beaulieu et al. 2006; Gould et al. 2006; Gaudi et al. 2008; Bennett et al. 2008) refer to quite obvious signatures. However, more subtle deviations are known to be frequent in the sample of observed events – of which none should be regarded as due to an isolated single lens star (Dominik & Hirshfeld 1996b; Di Stefano & Perna 1997) –, and these therefore significantly contribute to the statistics and must not be neglected. Less prominent features are associated with a larger degree of ambiguities and degeneracies, of which some are well-known (e.g. Mao & Di Stefano 1995; Dominik & Hirshfeld 1996a; Gaudi & Gould 1997; Dominik 1999b; Gaudi 1998; Gaudi & Han 2004) and involve planetary systems, stellar/substellar lens or source binaries, or the finite angular size of the source star. Ambiguities might be resolved in some cases by means of further observations identifying the lens star (Bennett et al. 2007) or obtaining an astrometric signature (Gould & Han 2000; Han 2002).

For the data acquired on any given event, the achieved planet detection efficiency is a natural function of the separation parameter $d$ and the mass ratio $q$ (Gaudi & Sackett 2000), two dimensionless parameters that characterize the observed planetary signal. For the vast majority of events, we cannot extract the mass $M$ of the lens star, and thereby the mass of a planet $m = q\,M$, but are only left with a measurement of the event time-scale $t_\mathrm{E} \equiv \theta_\mathrm{E}/\mu$, which also involves the relative proper motion $\mu$ between lens and source star as well as the relative parallax $\pi_\mathrm{LS} = 1\ \mathrm{AU}\,(D_\mathrm{L}^{-1} - D_\mathrm{S}^{-1})$. Similarly, instead of obtaining the orbital semi-major axis $a$, the related parameter $d$ just provides us with the current angular separation at unknown phase of an orbit with unknown inclination and eccentricity in units of the angular Einstein radius $\theta_\mathrm{E}$. In those cases, the best knowledge about these quantities is given by probability densities derived by means of Bayes' theorem with assumed priors for the mass densities and velocities of the source and lens stars as well as a lens star mass function (Dominik 2006). Using such probability densities, planet detection efficiencies as a function of the planet mass $m$ and the orbital semi-major axis $a$ can be obtained. There are however cases where the microlensing parallax $\pi_\mathrm{E} = \pi_\mathrm{LS}/\theta_\mathrm{E}$ significantly perturbs the light curve, and thereby can be determined by the acquired data (Alcock et al. 1995b; Hardy & Walker 1995), or where sensitivity to the finite angular size $\theta_\star$ of the source star significantly affects it by means of the parameter $\rho_\star = \theta_\star/\theta_\mathrm{E}$ (Witt & Mao 1994), so that a measurement of the angular Einstein radius $\theta_\mathrm{E}$ is being provided. In fact, the latter is the case for many events with planetary signals and almost certainly for those involving planets below $20\ M_\oplus$. Moreover, the relative proper motion $\mu$ can be measured if the lens star can be identified some time after the microlensing event took place, e.g. by means of HST observations (Bennett et al. 2007). Determination of any two of the parameters $\pi_\mathrm{E}$, $\theta_\mathrm{E}$, or $\mu$ yields the masses $m$ and $M$ of the planet and its host star with reasonable accuracy ($\sim$ 15–20 %), whereas the orbital semi-major axis $a$ still remains substantially uncertain due to the lack of any information about orbital phase, inclination, and eccentricity.

For each of the observed events and their respective model parameters, models of the Milky Way allow to determine a probability for the lens star to belong to either the Galactic disk ($\sim 1/3$ of the events) or the bulge ($\sim 2/3$ of the cases); and thereby the distributions of planets orbiting disk or bulge stars can be stochastically separated (Dominik 2006).

## 5 ONGOING EFFORTS AND CHALLENGES

### 5.1 Traditional approaches

The efforts undertaken by the OGLE (*O*ptical *G*ravitational *L*ensing *E*xperiment) and MOA (*M*icrolensing *O*bservations in *A*strophysics) surveys (Udalski et al. 1992; Muraki et al. 1999), monitoring $\gtrsim 2 \times 10^8$ Galactic bulge stars, lead to the identification of nearly 1000 ongoing microlensing events per year, where automated alert systems (Udalski et al. 1994; Bond et al.

2001) and the provision of real-time data[6 7] give follow-up campaigns the opportunity to choose from about 80 ongoing events at any time. Given that there is no single optimal global strategy for selecting the possible targets for further monitoring, and one can define different science goals, the current campaigns have adopted different approaches that match the capabilities of their respective instruments.

PLANET (*P*robing *L*ensing *A*nomalies *NET*work) went operational in 1995 with a network of 1m-class (semi-)dedicated optical telescopes that allow for a round-the-clock coverage (Albrow et al. 1998).[8] The desired photometric accuracy of 1–2 % and a sampling interval of 1.5 to 2.5 hrs, sufficient to characterize deviations by Jupiter-like planets, allows to monitor $\sim 20$ events on giant stars in the Galactic bulge or $\sim 6$ on main-sequence stars each night (Dominik et al. 2002).

Only with the OGLE-III upgrade in 2002, a sufficient number of microlensing events on bright targets to allow PLANET to get near its full capabilities were reported. Further upgrades to the surveys, with the OGLE-IV phase coming as soon as 2009, will see the number of detected microlensing events rising. This makes it well worth thinking about substantial improvements on the capabilities of follow-up campaigns, realized e.g. by using 2m or clusters of 1m telescopes.

Beginning with the deployment of the first of the used 2m robotic telescopes in 2004, RoboNet-1.0 (Burgdorf et al. 2007) has been exploiting a gradually rolled-out network of finally three such instruments, namely the Liverpool Telescope and the two Faulkes Telescopes, since 2005 as part of a common PLANET/RoboNet campaign. For 2008, the control over the two Faulkes Telescopes switched over to their new owner, the Las Cumbres Observatory Global Telescope Network (LCOGT.net), but the microlensing programme is continued as RoboNet-II in a new partnership (Tsapras 2008).

In contrast to PLANET and RoboNet, the MicroFUN (*Micro*lensing *F*ollow-*U*p *N*etwork) team (Gould et al. 2006) does not have quasi-continuous access to a network of telescopes. While a single site is used to target a few particularly promising events, a huge number of further sites can be activated on a target-of-opportunity basis. As pointed out in Sect. 3, such an operational setup is well-suited to hunt for planetary deviations near the predictable peak of microlensing events, and convinces by its good return-to-investment ratio. The extraordinary brightness of the highly-magnified targets near the peaks make them suitable for monitoring by amateur astronomers with 0.3m telescopes, some of which in association with MicroFUN already regularly make valuable contributions to forefront science. The much larger number of potential observing sites makes the MicroFUN network less vulnerable to weather losses than that of PLANET/RoboNet, which can turn out to be crucial.

### 5.2 Need for automation and its realization

The use of robotic telescopes for microlensing follow-up observations, pioneered by RoboNet-1.0, brought along the demand to deploy an automated algorithm for target selection and prioritization. Distributing the available time during the coming night for the specific characteristics of a specified telescope, web-PLOP (Snodgrass et al. 2008) is now freely available to the scientific community (or anyone else) and operates by means of a simple web-based interface[9].

[6] OGLE: `http://ogle.astrouw.edu.pl/ogle3/ews/ews.html`
[7] MOA: `https://it019909.massey.ac.nz/moa/alert/alert.php`
[8] Amongst various contributions, current PLANET operations are supported by ANR project "HOLMES".
[9] `http://www.artemis-uk.org/web-PLOP`

The world-leading software technology developed in order to match the RoboNet and PLANET demands for implementing a powerful strategy as well as controlling and visualizing the results has now been bundled into the ARTEMiS (*A*utomated *R*obotic *T*errestrial *E*xoplanet *Mi*crolensing *S*earch) expert system (Dominik et al. 2008), which enables a world-wide cooperative effort to hunt for extra-solar planets of Earth mass and below by microlensing, bringing together the collaborations acquiring the data and fostering communication not only between the involved scientists but also with the general public. An integral part of ARTEMiS is the SIGNALMEN anomaly detector (Dominik et al. 2007), which allows a prompt identification of potential planetary signatures and the immediate issue of requests to collect further data in order to confirm and characterize a suspected anomaly or to reject it.

Thumb estimates based on the number of monitored events and on 'typical' detection efficiencies (Gould & Loeb 1992; Bennett & Rhie 1996; Sackett 1997; Griest & Safizadeh 1998; Albrow et al. 2001; Gaudi et al. 2002) indicate an abundance around $\sim 2\,\%$ for planets with masses between 1 and 5 $M_\mathrm{jup}$ orbiting M dwarfs at radii between 1 and 4 AU, in agreement with the findings from radial-velocity surveys (e.g. Udry & Santos 2007; Santos 2008), whereas the abundance of planets between 2 and 10 $M_\oplus$ in the same orbital range is expected to be between 60 % and 200 %.

Given that low-mass planets seem to be common around M dwarfs, it is straightforward to equip microlensing planet searches with the capabilities to explore the super-Earth mass regime, and move towards the detection of Earth-mass planets (Bennett & Rhie 1996). As shown in Fig. 6, a cooperative three-step strategy of survey, follow-up, and anomaly monitoring, making use of an automated anomaly detector with immediate feedback, such as SIGNALMEN (Dominik et al. 2007), provides current or near-future facilities with a realistic opportunity for this to happen. Namely, if one manages to monitor 100 events per observing season with a sampling interval $\Delta t = (90~\mathrm{min})\,A^{-1/2}$, one would expect to detect one planet between 0.5 and 2 $M_\oplus$ per year, assuming every star hosts one of these on average at orbital radii between 1 and 4 AU, and the respective planet detection efficiency for the observed events averages 1 %.

As illustrated in Fig. 7, microlensing planet searches are moreover in principle capable of breaking the Earth-mass barrier. While the reduction of the signal amplitude tenders already the detection of Super-Earths impossible if a giant source star gets very closely aligned with the orbited lens star, such targets are still favourable for detecting Earth-mass planets at moderate magnifications (i.e. larger angular separations). For substantially less massive planets however, one is forced to rely on main-sequence source stars (Han 2007a).

Robotic Telescopes are not only cost-efficient by allowing human resources to be freed from time-consuming low-profile tasks and to be concentrated on the science itself, but are also capable of flexible scheduling and immediate response. The chances for detecting planets of Earth mass and below critically depend on the news of a potential planetary signal being observable to be spread and further data being taken within 5–10 min. Achieving such a result with the vast amount of data that needs to be assessed immediately requires an automated approach.

Gravitational microlensing acts as a major science driver for the development of the infrastructure and technology for realizing time-domain astronomy. For scheduling its observations, RoboNet-1.0 already used the novel software architecture developed by the eSTAR (*e*-*S*cience *T*elescopes for *A*stronomical

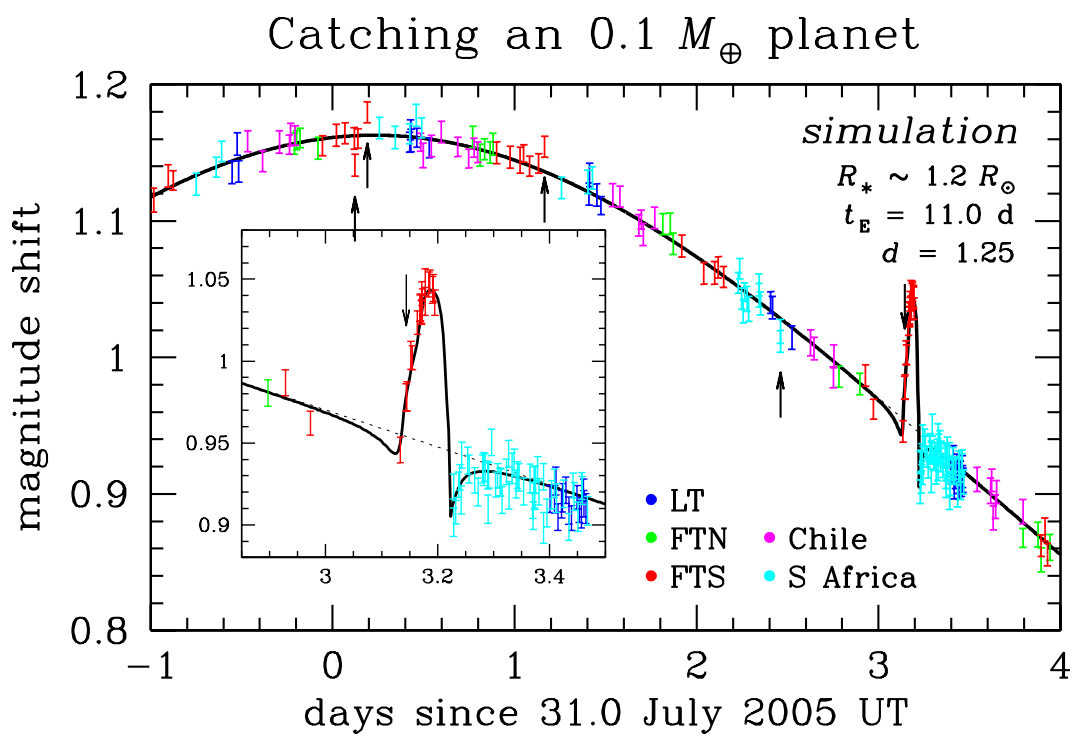

**Figure 7.** Simulation of the detection of a planet with 0.1 Earth masses using a three-step strategy of survey, follow-up, and anomaly monitoring. The adopted parameters resemble those found for event OGLE-2005-BLG-390 (Beaulieu et al. 2006), where the closest angular approach $u_0 = 0.359$ between lens and source star occurs at epoch $t_0 = 31.231$ July 2005 UT and the event time-scale is $t_\mathrm{E} = 11.0$ d. Differing from this event, the planet is 55 times less massive ($q = 1.5 \times 10^{-6}$), corresponding to $m \sim 0.1\ M_\oplus$, the observed target is a main-sequence star ($R_\star \sim 1.2\ R_\odot$) instead of an 8 times larger giant, so that $\rho_\star \equiv R_\star/(D_\mathrm{S}\,\theta_\mathrm{E}) = 0.0032$, the angular separation of the planet from its host star has been shifted to the more favourable $d = 1.25$ times the angular Einstein radius $\theta_\mathrm{E}$, and the angle of the source trajectory with respect to the planet-to-star axis has been changed to $\alpha = 126.6°$, so that the source passes over the planetary caustic. The photometric accuracy has assumed to be 1 % for the unbrightened source star, and less as it brightens (following photon noise), where a further systematic error of 0.5 % has been added in quadrature and a fluctuation of 12.5 % has been adopted. Such a precision is achievable for well-isolated targets (in fact, there are real data actually being better), but given the crowding level of the Galactic bulge, such do not constitute the majority. Nevertheless, the signal would still be clearly detectable with accuracies $\lesssim 3\,\%$. The average sampling interval is $\Delta t = (2\ h)/\sqrt{A}$ with a 20 % fluctuation, where $A$ denotes the current magnification. The adopted network involves the 2m robotic telescopes currently exploited by RoboNet (Burgdorf et al. 2007; Tsapras 2008), augmented by two further hypothetical identical ones in South Africa and Chile. The black arrows show the points on which the automated anomaly detector (Dominik et al. 2007) triggered. While this figure shows the opportunities, it also demonstrates three key challenges: achieving an accurate and reliable real-time photometry, assessing the collected data and disseminating the result within minutes, and having a telescope at a suitable location ready to observe. In the case shown, the planetary anomaly was detected, but a 40 min gap between Eastern Australia and South Africa turned out to be too long for being able to acquire enough data for properly characterizing it.

*R*esearch) project (Steele et al. 2002), which builds a virtual meta-network between existing proprietary robotic-telescope networks providing a uniform interface based on a multi-agent contract model (Allan et al. 2006) and the exchange of standardized messages in RTML (*R*emote *T*elescope *M*arkup *L*anguage) (Hessman 2001; Pennypacker et al. 2002; Hessman 2006). Immediate response is being realized by means of VOEvents (*V*irtual *O*bservatory *Events*) (Williams & Seaman 2006), a standard maintained by the IVOA (*I*nternational *V*irtual *O*bservatory *A*lliance) (Quinn et al. 2004), which are also made available as RSS 2.0 (*R*eally *S*imple *S*yndication) feeds. Information about the OGLE and MOA microlensing alerts is already available this form, and once ARTEMiS complies with these standards, it will immediately be linked to all robotic telescopes of the HTN (*H*eterogeneous *T*elescope *N*etworks) consortium, which includes several sites with a declared interest in microlensing observations, namely the LCOGT.net telescopes (with the network to be vastly expanded from 2009 onwards) and the Liverpool Telescope – both used now by RoboNet-II – as well as the two MONET (*MO*nitoring *NE*twork of *T*elescopes) telescopes (Hessman 2004).

Further-reaching opportunities will be provided by the upcoming SONG (*S*tellar *O*bservations *N*etwork *G*roup) telescopes, placed at 8 locations spread in longitude over both hemispheres, with a prototype being commissioned in early 2009. The individual 1m telescopes, to be clustered, are being designed to reach the diffraction limit in $I$-band for that diameter. Together with further advanced telescope technology, such as a high-speed read-out, these will allow to go three magnitudes deeper than possible with the instruments currently being used (Jørgensen 2008). With these properties, the SONG network becomes strongly competitive with a space mission (that cannot achieve a better resolution for the same telescope size), in particular given that its lifetime will be substantially larger, and servicing as well as upgrades can be done easily. The trade-off for the impressive capabilities is in the extremely narrow field-of-view, so that for a microlensing programme to be successful, linking up with a matching survey is required.

During 2008, ARTEMiS intends to fully develop an advanced target selection algorithm providing a recommendation that follows the defined preferences, commitments and aims of the observing campaign that owns time at the considered telescope. It will therefore have to consider carefully four different groups of input: the capabilities of the hardware infrastructure, the strategy followed at each observing site as defined by specific science goals, the full set of data available at present, and the current observability (Dominik 2008).

MiNDSTEp (*Mi*crolensing *N*etwork for the *D*etection of *S*mall *T*errestrial *Ep*oplanets)[10], emerging from the experience gained with PLANET and advancing further, is the prototype for Generation-IV microlensing follow-up, and in particular develops the operational context for the microlensing programme to be deployed on the SONG telescopes. The MiNDSTEp campaign adopts a fully-deterministic ARTEMiS-assisted strategy, which allows a proper assessment of the abundance of low-mass planets within the Milky Way by means of Monte-Carlo simulation, overcoming any contamination of the statistics by unpredictable and irreproducible human judgement.

### 5.3 Next-generation event modelling

The lack of an automated modelling system for anomalous microlensing events already poses a severe bottleneck for the ongoing data analysis. With increased capabilities in sight, the current human-modeller approach will become unsustainable. Moreover, among the about 80 anomalous events per season, only a few are proper candidates for a planetary nature of the observed deviations. If all detected anomalies are densely monitored over prolonged times, a lot of efforts are wasted and insufficient time is spent on the regular monitoring of events that has to deliver the anomaly candidates. Therefore, a real-time assessment of ongoing anomalies is strongly desired.

There is a rather straightforward procedure for determining the model parameters characterizing a planetary deviation that results from the source getting close to planetary caustics after it has been observed over its full course (Gaudi & Gould 1997), while this holds to a lesser extent for approaches to the central caustic (Han et al. 2001). However, in the early stages of an ongoing anomaly, one is plagued with ambiguities and degeneracies that appear to prevent a timely assessment of the mass ratio of a supposed binary lens and therefore e.g. prevents to decide whether one sees the effect of a binary star or of a planet orbiting a star. While a first characteristic feature that allows a discrimination has been identified recently (Han & Gaudi 2008), this issue needs some further investigation.

In order to take care of the intricate high-dimensional parameter space, various approaches have been suggested, such as an event library for initial parameter seeds (Mao & Di Stefano 1995), super-cluster computing (Rattenbury et al. 2002), and the use of genetic algorithms (Kubas 2005). Artificial neural networks could map the decision processes carried out in the brains of the best-trained scientists that are currently kept busy with this task. While blind approaches have shown to be everything but powerful, the combination with an optimal characterization of the observed features appears to be promising (Vermaak 2007), and in fact the finite number of topologies of the caustics and the possible relative positions of the source trajectory lead to a finite set of morphologies for microlensing light curves due to stellar binaries and planetary systems. A related classification with regard to the images has already been discussed by Bozza (2001). Step-by-step manual approaches that identify the characteristic features of an observed light curve and map these to trial solutions have proven successful (Dominik & Hirshfeld 1996a; Dominik 1999a; Albrow et al. 1999).

[10] `http://www.mindstep-science.org`

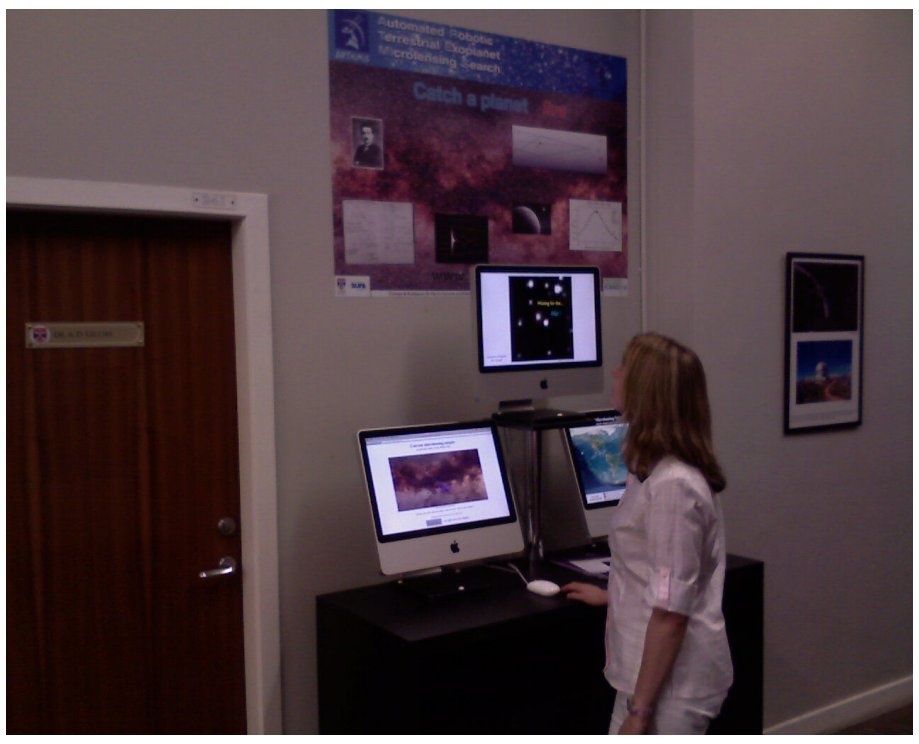

**Figure 8.** "Catch-a-planet" microlensing display powered by ARTEMiS, which shows the current network status along with visualized live data, and provides background information.

### 5.4 Engaging the public

Scientific research on a substantial scale can only be carried out in first instance because it has the general support of the society. Moreover, the genuine role of a scientist is to increase the knowledge of the society (rather than his/her own). Therefore, communication needs to constitute an essential and integral part of "scientific" work. ARTEMiS is therefore committed to not only provide powerful tools to scientists, but also to engage in a dialogue with the general public about the detection of new worlds, thereby sharing the enthusiasm and vision. A "Catch-a-planet" display providing current model light curves of monitored events along with live data (some of these appearing just minutes after the observations were taken) was on show at the 2008 Royal Society Summer Science Exhibition[11] as part of an exhibit entitled "Is there anybody out there? Looking for new worlds", and is now semi-permanently installed in the foyer of the Physics & Astronomy building of the University of St Andrews (see Fig. 8). In a further initiative, there are plans to involve schools in the observations themselves via LCOGT.net and the Faulkes Telescopes Project[12]. The detection of extra-solar planets (not only by gravitational microlensing) as well as the search for extra-terrestrial life constitute thankful topics for getting in touch with the general public, and one should not miss out on the potential that this harbours for the support of our scientific research.

## 6 SOME FUTURE OPPORTUNITIES

Provided that envisioned projects are adequately funded, capabilities of microlensing planet searches will increase both by employing advanced technology and by scaling up facilities: the OGLE-IV upgrade, the extension of robotic-telescope networks (Tsapras 2008), high-cadence wide-field surveys (Han 2007b), the SONG diffraction-limited telescopes (Jørgensen 2008), and a possible microlensing space mission (Bennett & Rhie 2002; Bennett et al. 2003) have already been mentioned.

Apart from a brightening, gravitational microlensing also leads to an apparent positional shift of the observed target, proportional to the angular Einstein radius $\theta_\mathrm{E}$, which can be measured from such an astrometric signature (Høg et al. 1995; Miyamoto & Yoshii 1995; Walker 1995), coming in different flavours (Dominik & Sahu 2000). Astrometric microlensing has been hailed as an exciting alternative channel to detect planets (Safizadeh et al. 1999), working quite well for gas giants, but disfavouring less massive planets in proportion to the planet-to-star mass ratio $q$ rather than $\sqrt{q}$ as for photometric microlensing (Dominik 2001a; Han & Lee 2002). In any case, astrometric observations provide further valuable information about ongoing microlensing events that have been detected from a photometric signature, and in particular resolve all known generic ambiguities (Gould & Han 2000; Han 2002). Contrary to expectations in the late 1990s, the desired micro-arcsecond astrometry on specified targets has not become available due to SIM (*S*pace *I*nterferometry *M*ission) not going into operation. However, the combination of a list of ongoing microlensing events – automatically created by systems like ARTEMiS – with the pre-programmed observing schedule of ESA's GAIA mission might well result in identifying a few but crucial high-precision astrometric measurements after its launch date set for December 2011. Access to suitable ground-based interferometry (such as the VLTI) is extremely limited, so that something would need to be done about available facilities for making astrometric microlensing observations a generic useful tool. Adopting the noise-filtration technique proposed by Lazorenko & Lazorenko (2004), which has been tested successfully (Lazorenko 2006; Lazorenko et al. 2007), could offer a path to achieve a high accuracy with a single large telescope.

Microlensing is not only able to reveal planets orbiting distant stars within the Milky Way, but its potential even extends to studying the abundance of planets in neighbouring galaxies such as M31 (Covone et al. 2000). With unresolved targets, the sensitivity is almost exclusively restricted to gas-giant planets and events on giant source stars during high-magnification phases. Recent efforts have seen the implementation of a real-time alert system on ongoing events (Darnley et al. 2007), marking a milestone towards building a network with alert-driven follow-up capabilities that allow to detect planetary signals (Dominik 2001b; Chung et al. 2006).

Apart from revealing planets orbiting the lens star by means of a photometric or astrometric signal, some further suggestions that rely on a microlensing signature include detecting planets around source stars during caustic passages (Graff & Gaudi 2000; Gaudi et al. 2003) or due to the stellar reflex orbital motion around the common barycentre (Rahvar & Dominik 2008).

The technique of gravitational microlensing has developed quite a lot over the last 15 years, and it is probably not too wrong to expect that it will further progress a lot over the next such time-span. Some of the advances and opportunities to come might be as surprising, sudden, and unforeseeable to us now as part of the current achievements were 15 years ago. Not even Einstein could have imagined how some decades of advance in technology would revolutionize the potential of applications of a simple physical effect that he once judged upon (Einstein 1936): "...there is no great chance of observing this phenomenon ...".

[11] http://www.summerscience.org.uk
[12] http://www.faulkes-telescopes.com

## IMAGE CREDITS

*Figure1*: ESO, NASA, ESA, A. Schaller (STScI), G. Bacon (STScI), Korean Astronomy and Space Science Institute (KASI), Chungbuk National University, Astrophysical Research Center for the Structure and Evolution of the Cosmos (ARCSEC) – (respective parts)
*Figure2*: "Atlas Image mosaic obtained as part of the Two Micron All Sky Survey (2MASS), a joint project of the University of Massachusetts and the Infrared Processing and Analysis Center/California Institute of Technology, funded by the National Aeronautics and Space Administration and the National Science Foundation." – (part)

## ACKNOWLEDGMENTS

We would like to thank D. Lin and S. Ida for advice, suggestions, and encouragement.